\documentclass{elsart}
\usepackage{epsfig}

\def\Shat{\hat{s}}
\def\Q2tilde{\tilde{Q}^2}
\def\MSbar{$\overline{\mathrm{MS}}\ $}
\def\msbar{\overline{\tiny \mathrm{MS}}}
\def\Li#1#2{{\mathrm{Li}}_{#1}\left(#2\right)}

\def\sww{\sin^2\theta_{\sss W}}
\def\ba{\begin{eqnarray}}
\def\ea{\end{eqnarray}}
\def\dd{{\mathrm d}}

\def\fun#1#2{\lower3.6pt\vbox{\baselineskip0pt\lineskip.9pt
  \ialign{$\mathsurround=0pt#1\hfil##\hfil$\crcr#2\crcr\sim\crcr}}}
\def\order#1{{\mathcal O}\left(#1\right)}

\newcommand{\nll}{\nonumber\\}
\newcommand{\sss}[1]{\scriptscriptstyle{#1}}
\def\mw {M_{\sss W}}
\def\mz {M_{\sss Z}}
\def\mh {M_{\sss H}}
\def\stw{s_{\sss W}}
\def\ctw{c_{\sss W}}

\begin{document}

\begin{frontmatter}

\title{Radiative Corrections to Neutrino Deep Inelastic Scattering Revisited}

\author[BLTP]{A.B. Arbuzov}, 
\author[LYaP]{D.Yu. Bardin}, and 
\author[LYaP]{L.V. Kalinovskaya}

\address[BLTP]{Bogoliubov Laboratory of Theoretical Physics, \\
JINR,\ Dubna, \ 141980 \ \  Russia}
\address[LYaP]{Dzhelepov Laboratory of Nuclear Problems, \\
JINR,\ Dubna, \ 141980 \ \  Russia}

\begin{abstract}
Radiative corrections to neutrino deep inelastic scattering are revisited.
One--loop electroweak corrections are re--calculated within the automatic SANC 
system. Terms with mass singularities are
treated including higher order leading logarithmic corrections.
Scheme dependence of corrections due to weak interactions is investigated.
The results are implemented into the data analysis of the NOMAD experiment.
The present theoretical accuracy in description of the process is discussed. 
\end{abstract}


\end{frontmatter}

\section{Introduction}

Modern experiments such as NOMAD~\cite{Astier:2003rj},
NuTeV~\cite{McFarland:gx}, 
and CHORUS~\cite{Kayis-Topaksu:ds}
made a serious step forward in studies of neutrino deep inelastic scattering. 
Their precision measurements made it necessary to update
the accuracy level of the theoretical description of the process. 
Important ingredients for an advanced precision in the theoretical 
predictions is the calculation of the relevant radiative corrections (RC).

Our study is motivated by the request from the NOMAD experiment. 
Electroweak (EW) radiative corrections to neutrino--nucleon scattering 
should have been implemented into the general Monte Carlo system for the
experimental data analysis. Certain experimental conditions of particle
registration and event selection should have been taken into account.
In order to make the relevant subroutine describing the corrections 
fast, we had to look for an analytical answer (without numerical 
integrations). 

In the present work we reproduce most of the results of unpublished
communication~\cite{Bardin:1986bc}. Contrary to the earlier calculation, we
used the modern technique of automatic calculations within the
SANC system~\cite{Andonov:2002jg,Bardin:zd,SANCwww,Andonov:2002xc} 
developed for Support
of Analytic and Numeric calculations for experiments at Colliders.
Besides the one-loop calculation, we consider certain contributions
of higher orders and discuss the theoretical uncertainty
due to unknown electroweak (EW) corrections for the case of 
the concrete experimental study.  

The paper is organized as follows. In the next section we define
the notation and present the Born level distributions. Then we consider
different sources of radiative corrections. Special subsections are
devoted to quark and muon mass singularities. Numerical results and
comparisons are presented in Sect.~\ref{NumRes}. Possible applications
of our formulae are discussed in Conclusions.

\section{The Born Cross Section}

We will consider the process of deep inelastic scattering (DIS) 
in the framework of the
quark--parton model, assuming that we are in the proper kinematical region
$(Q^2 \gg \Lambda_{QCD}^2)$. 

Here we list the Born level cross sections of neutrino--quark interaction
weighted by the quark density functions.
For the charge current (CC) scattering processes 
\ba
\nu(k_1) + q_i(p_1) \to l^-(k_2) + q_f(p_2)  
\  \   \mathrm{and} \  \
\bar{\nu} + q_i  \to l^+ + q_f
\ea
we have 
\ba \label{SnuqCC}
&& \frac{\dd^2\sigma_{\nu  CC}^{\mathrm{Born}}}{\dd x\dd y} = \sigma^0_{CC},
\qquad
\frac{\dd^2\sigma_{\bar{\nu} CC}^{\mathrm{Born}}}{\dd x\dd y} = \sigma^0_{CC}(1-y)^2,
\\
&& \sigma^0_{CC} = |V_{if}|^2\frac{G_F^2}{\pi}\hat{s}
\frac{M_{\sss W}^4}{(M_{\sss W}^2+\hat{Q}^2)^2}f_i(x,\hat{Q}^2),
\ea
where $M_{\sss W}$ is the $W$-boson mass; $\hat{s}$ is the center--of--mass energy of
the neutrino-quark system squared; $|V_{if}|$ is the element of 
the Cabibbo-Kobayashi-Maskawa quark 
mixing matrix; $f_i(x,\hat{Q}^2)$ is the density function of the initial 
quark in the given nucleon. 
The kinematics\footnote{We use the $(+,-,-,-)$ metrics, $p=(p^0,\vec{p}).$} 
is described by the Bjorken variables:
\ba
&& y=\frac{\hat{Q}^2}{\hat{s}}, \qquad
\hat{Q}^2 = -(p_2-p_1)^2, \qquad
x=\frac{\hat{Q}^2}{yS}, 
\nonumber \\
&& \hat{s} = (k_1+p_1)^2 \approx xS, \qquad
S = (k_1 + P)^2,
\ea
where $P$ is the initial nucleon momentum.
We will use $m_{1,2,l}$  $(Q_{1,2,l})$ for the masses (charges) of the initial 
quark, the final quark, and the muon. 
 
For the neutral current (NC) scattering processes
\ba
\nu(k_1) + q_i(p_1) \to \nu(k_2) + q_f(p_2) 
\  \   \mathrm{and} \  \
\bar{\nu} + q_i  \to \bar{\nu} + q_f
\ea
we have
\ba \label{SnuqNC}
&& \frac{\dd^2\sigma_{\nu NC}^{\mathrm{Born}}}{\dd x\dd y} = \sigma^0_{NC}\biggl(
g_L^2 + g_R^2(1-y)^2 - \frac{2m_1^2y}{\hat{s}}g_L g_R\biggr), 
\\
&& \frac{\dd^2\sigma_{\bar{\nu}  NC}^{\mathrm{Born}}}{\dd x\dd y} = \sigma^0_{NC}\biggl(
g_R^2 + g_L^2(1-y)^2 - \frac{2m_1^2y}{\hat{s}}g_L g_R\biggr), 
\\
&& \sigma^0_{NC} = \frac{G_F^2}{\pi}\hat{s}\frac{M_{\sss Z}^4}{(M_{\sss Z}^2
+\hat{Q}^2)^2}f_i(x,\hat{Q}^2),
\ea
where
\ba
g_L = - \frac{1}{2} + |Q_i|\sww, \qquad g_R = |Q_i|\sww, 
\ea
$M_{\sss Z}$ is the $Z$-boson mass; and $\theta_{\sss W}$ is the weak mixing angle.

\section{Radiative Corrections}

The radiatively corrected neutrino DIS cross section can be represented as the sum 
of the Born distribution with the 
contributions due to virtual loop diagrams (Virt), soft photon emission (Soft),
and hard photon emission (Hard):
\ba
\frac{\dd^2\sigma_{i}^{\mathrm{Corr.}}}{\dd x\;\dd y} = 
\frac{\dd^2\sigma_{i}^{\mathrm{Born}}}{\dd x\;\dd y}
\biggl( 1 + \delta_i^{\mathrm{Virt}}  + \delta_i^{\mathrm{Soft}}  
+ \delta_i^{\mathrm{Hard}} \biggr),   
\ea
where the index $i$ denotes the type of the process under consideration ($\nu q\; CC$,
$\bar{\nu}q\; NC$ and so on). This formula assumes calculation of the order 
$\order{\alpha}$ RC. 
Certain higher order corrections will be added below as well.
We assume also that the momentum transfer
square is small compared with the $W$-boson mass: corrections of the order
$\alpha Q^2/M_{\sss W}^2$ are omitted from the calculation.

\subsection{Virtual Corrections}

Contributions due to virtual (one--loop) EW corrections are
re--calculated by means of the SANC system.
All the contributing one--loop diagrams were calculated in the frame of
the Standard Model. Omitting the terms suppressed by the 
$Q^2/M_{\sss W}^2$ ratio allowed us to get the short analytical answer. 

The virtual correction to the CC neutrino--quark scattering reads
\ba
\delta^{\mathrm{Virt}}_{\nu CC} &=& \frac{\alpha}{\pi}\biggl\{
      -\Biggl[ \frac{1}{2} Q_1^2     \ln\frac{\hat{Q}^2}{m^2_1}
        + \frac{1}{2}(1 + Q_1)^2     \ln\frac{\hat{s}}{m^2_2}     
        + \frac{1}{2}                \ln\frac{\hat{s}}{m^2_l}        
\nll &&
        + \left(Q_1+\frac{Q_1^2}{2}\right)\ln y 
        - Q_1                        \ln(1 - y) 
        - 1 - Q_1 - Q_1^2  \Biggr]        \ln\frac{\hat{s}}{\lambda^2}
\nll &&
        +\frac{1}{4}\Biggl[ Q_1^2    \ln^2\frac{\hat{Q}^2}{m^2_1}  
        + (1 + Q_1)^2                \ln^2\frac{\hat{s}}{m^2_2}
        +                            \ln^2\frac{\hat{s}}{m^2_l}
        +                            \ln\frac{\hat{s}}{m^2_l}
\nll &&
        + Q_1^2 ( 1 - 2\ln y )       \ln\frac{\hat{Q}^2}{m^2_1}       
        + (1 + Q_1)^2                \ln\frac{\hat{s}}{m^2_2} 
        \Biggr]
\nll &&
        - \frac{1}{4}\Biggl[  Q_1(2 + Q_1)     \ln^2 y 
                            - 2 Q_1            \ln^2 (1 - y)  
                            - Q_1 (6 + 5 Q_1)  \ln y  
\nll &&
                            + 6 Q_1            \ln(1 - y) 
                            - 4 (2 + Q_1)^2    \zeta_2       
                            + 1 + Q_1 + 8 Q_1^2\Biggr]     
\nll &&
        - \frac{3}{2} ( 1 + Q_1 ) \ln\frac{\hat{s}}{M_{\sss Z}^2}\biggr\},
\\ \nonumber &&
\zeta_2 = \Li{2}{1} = \frac{\pi^2}{6}, \qquad 
\Li{2}{x} = - \int\limits_{0}^{1}\frac{\ln(1-xy)}{y}\dd y,
\ea
where $\lambda$ is the auxiliary photon mass, $\lambda \ll m_{1,2,l}$.
Note that in the above expression we used the explicit value $Q_l=-1$
and eliminated the final state quark charge by applying the charge conservation law,
$Q_2 = Q_1 - Q_l$.

For the CC antineutrino--quark scattering we get
\ba
\delta^{\mathrm{Virt}}_{\bar{\nu} CC} &=& \frac{\alpha}{\pi}\biggl\{
      -\Biggl[ \frac{1}{2} Q_1^2     \ln\frac{\hat{Q}^2}{m^2_1}
        + \frac{1}{2}(1 - Q_1)^2     \ln\frac{\hat{s}}{m^2_2}     
        + \frac{1}{2}                \ln\frac{\hat{s}}{m^2_l}        
\nll &&
        - \left(Q_1-\frac{Q_1^2}{2}\right)\ln y 
        + Q_1                        \ln(1 - y) 
        - 1 + Q_1 - Q_1^2  \Biggr]        \ln\frac{\hat{s}}{\lambda^2}
\nll &&
        +\frac{1}{4}\Biggl[ Q_1^2    \ln^2\frac{\hat{Q}^2}{m^2_1}  
        + (1 - Q_1)^2                \ln^2\frac{\hat{s}}{m^2_2}
        +                            \ln^2\frac{\hat{s}}{m^2_l}
        +                            \ln\frac{\hat{s}}{m^2_l}
\nll &&
        + Q_1^2 ( 1 - 2\ln y )       \ln\frac{\hat{Q}^2}{m^2_1}       
        + (1 - Q_1)^2                \ln\frac{\hat{s}}{m^2_2} 
        \Biggr]
\nll &&
        + \frac{1}{4}\Biggl[  Q_1(2 - Q_1)     \ln^2 y 
                            - 2 Q_1            \ln^2 (1 - y)  
                            - Q_1 (6 - 5 Q_1)  \ln y  
\nll &&
                            + 4 (4 - Q_1 + Q_1^2)    \zeta_2       
                            - 8 + 15Q_1 - 8 Q_1^2\Biggr]     
        - \frac{3}{2} Q_1 \ln\frac{\hat{s}}{M_{\sss Z}^2}
\biggr\}.
\ea
For the pure QED part of the virtual corrections to the NC neutrino (and antineutrino) 
scattering we have
\ba
\delta^{\mathrm{Virt}}_{NC} &=& \frac{\alpha}{\pi}Q_1^2\biggl\{
-\frac{1}{2}\ln\frac{\hat{Q}^2}{\lambda^2}\biggl(\ln\frac{\hat{Q}^2}{m_1^2}
+ \ln\frac{\hat{Q}^2}{m_2^2} -2\biggr)+ \frac{1}{4}\ln^2\frac{\hat{Q}^2}{m_1^2} 
+ \frac{1}{4}\ln^2\frac{\hat{Q}^2}{m_2^2}
\nll &&
+ \frac{1}{4}\biggl( \ln\frac{\hat{Q}^2}{m_1^2} + \ln\frac{\hat{Q}^2}{m_2^2}\biggr)
- 2 + \zeta_2 \biggr\}.
\ea
The weak part of the virtual correction to the NC case is included in the definition
of the effective electroweak couplings:
\ba
g_{L(R)} \to \tilde{g}_{L(R)}, \quad
\tilde{g}_L = \rho\biggl(- \frac{1}{2} +  |Q_i|\kappa\sww\biggr), \quad
\tilde{g}_R = \rho \kappa |Q_i|\sww,
\ea
where $\rho$ and $\kappa$ are the electroweak form factors,
\ba
\!\!\! \rho &=& \frac{3}{4}  \biggl\{
   \frac{1}{\stw^2}\ln\left( \frac{\mw^2}{\mz^2} \right)
  +\frac{\mh^2}{\mw^2}\left[ \frac{1}{(1-\mh^2/\mz^2)}\ln\left( \frac{\mh^2}{\mz^2} \right)
                            -\frac{1}{(1-\mh^2/\mw^2)}\ln\left( \frac{\mh^2}{\mw^2} \right)
                      \right]
\nonumber \\ && \!\!\!
             +1-4 I^{(3)}_f  
             -\frac{4}{\ctw^2} \left( \frac{1}{2} I^{(3)}_f-\stw^2 c_f
             +4 I^{(3)}_f \stw^4 c_f^2 \right)
             + \frac{m_t^2}{\mw^2} \biggr\},
\\
\!\!\! \kappa &=& 
              -\frac{\ctw^2}{\stw^2} \Delta\rho - \frac{7}{2}+3 I^{(3)}_f-\frac{2}{3} \ctw^2
              +\frac{3}{\ctw^2} \left( \frac{1}{2} I^{(3)}_f - \frac{3}{2} \stw^2 c_f
              +4 I^{(3)}_f \stw^4 c_f^2 \right)
\nonumber \\ && \!\!\!
              +2 B_{f}\big(Q^2;m_f,m_f\big)             
            -\Pi_{Z\gamma}^{\rm fer}(Q^2),
\ea
where $I^{(3)}_f$, $c_f$, $Q_f$, $v_f$ and $m_f$ are weak isospin, color factor 
(1 for leptons, 3 for quarks), charge, vector coupling and mass of a fermion; 
notation for $W,Z,H,t$ quark masses are evident; $\stw$ and $\ctw$ are
sine and cosine of weak mixing angle and $\stw^2=1-\mw^2/\mz^2$;
$\Delta\rho$ is Veltman parameter, 
\begin{equation}
\Delta\rho=\frac{1}{\mw^2}\left[\Sigma(\mw^2)-\Sigma(\mz^2)\right];
\end{equation}
function $B_{f}\big(Q^2;m_f,m_f\big)$ is a useful combination of 
the finite parts of the standard
Passarino--Veltman functions (without UV pole $1/\bar{\epsilon}$)
\begin{equation}
B_{f}\big(Q^2;m_f,m_f\big)=
2\left[B_{21}\big(Q^2;m_f,m_f\big)+B_{1}\big(Q^2;m_f,m_f\big)\right];
\end{equation}
and finally $\Pi_{Z\gamma}^{\rm fer}(Q^2)$ is the $\gamma Z$ mixing operator
\begin{equation}
\Pi_{Z\gamma}^{\rm fer}(Q^2) = 2 \sum_f c_f\,Q_f\,v_f\,B_{f}\big(Q^2;m_f,m_f\big).
\label{Pi_zg_fer}
\end{equation}
Note, that the three last quantities are taken at $\mu=\mw$, 
with $\mu$ being the t'Hooft scale.
More about the calculation and renormalization scheme 
applied here can be found in the book~\cite{BP:book}.

\subsection{Soft Photon Radiation}

Emission of a soft photon in neutrino DIS can be described in 
the standard way  by the accompanying radiation factors:
\ba
&& \delta^{\mathrm{Soft}}_{NC} = 
- Q_1^2\frac{\alpha}{4\pi^2}\int\frac{\dd^3k}{\omega}
\left( \frac{p_1}{p_1k} - \frac{p_2}{p_2k} \right)^2,
\\
&& \delta^{\mathrm{Soft}}_{CC} = - \frac{\alpha}{4\pi^2}
\int\frac{\dd^3k}{\omega}
\left( Q_1 \frac{p_1}{p_1k} - Q_2 \frac{p_2}{p_2k} - Q_l\frac{k_2}{k_2k}
\right)^2.
\ea

We consider the problem in the rest reference frame of the final quark,
$\vec{p}_2=0$, which is equivalent (in the soft photon limit) to the so--called 
$R$-reference frame,
used below in the evaluation of the hard photon contribution. 
The soft photon energy, $\omega$, is limited by the parameter $\bar\omega$, which
is assumed to be small compared to the large energy scale: $\bar\omega \ll \sqrt{Q^2}$.
List of the relevant integrals is given in Appendix~A. 
For the charged current we get
\ba \label{SoftCC}
&& \delta^{\mathrm{Soft}}_{\nu CC} = \frac{\alpha}{\pi}\Biggl\{
       \Biggl[ \frac{1}{2} Q_1^2     \ln\frac{\hat{Q}^2}{m^2_1}
        + \frac{1}{2}(1 + Q_1)^2     \ln\frac{\Shat}{m^2_2}
        + \frac{1}{2}                \ln\frac{\Shat}{m^2_l}
\nll && \quad
        + Q_1\left(1+\frac{1}{2}Q_1\right)\ln y  
        - Q_1                        \ln(1 - y) 
        - 1 - Q_1 - Q_1^2  \Biggr]        \ln\frac{\Shat}{\lambda^2}
\nll &&  \quad
     - \Biggl[ Q_1^2                 \ln\frac{\hat{Q}^2}{m^2_1}
        + ( 1 + Q_1)^2               \ln\frac{\Shat}{m^2_2}
        +                            \ln\frac{\Shat}{m^2_l}
\nll && \quad
        + Q_1 (2 + Q_1 )             \ln y                   
        - 2 Q_1                      \ln(1 - y)            
        - 2 (1 + Q_1 + Q_1^2 )\Biggr]\ln\frac{\Shat}{2\bar\omega m_2}
\nll && \quad
    - \frac{1}{4}\Biggl[ Q_1^2       \ln^2\frac{\hat{Q}^2}{m^2_1}
        - (1 + Q_1)^2                \ln^2\frac{\Shat}{m^2_2}
        +                            \ln^2\frac{\Shat}{m^2_l}
        -2 Q_1^2 (1 - \ln y)         \ln\frac{\hat{Q}^2}{m^2_1} 
\nll && \quad
        + 2 (1 + Q_1)^2              \ln\frac{\Shat}{m^2_2}
        - 2                          \ln\frac{\Shat}{m^2_l}\Biggr]
        - \frac{1}{4} Q_1 (2 + Q_1)  \ln^2 y 
        - \frac{1}{2} Q_1            \ln^2 (1 - y)  
\nll && \quad
        + Q_1                        \ln y \ln(1 - y)                      
        + \frac{1}{2} Q_1^2          \ln y         
        - ( 1 + Q_1 + Q_1^2 )        \zeta_2
        + ( 1 + Q_1 )^2
\Biggr\},
\\
&& \delta^{\mathrm{Soft}}_{\bar{\nu} CC} = \frac{\alpha}{\pi}\Biggl\{
       \Biggl[ \frac{1}{2} Q_1^2     \ln\frac{\hat{Q}^2}{m^2_1}
        + \frac{1}{2}(1 - Q_1)^2     \ln\frac{\Shat}{m^2_2}
        + \frac{1}{2}                \ln\frac{\Shat}{m^2_l}
\nll && \quad
        - Q_1\left(1-\frac{1}{2}Q_1\right)\ln y  
        + Q_1                        \ln(1 - y) 
        - 1 + Q_1 - Q_1^2  \Biggr]        \ln\frac{\Shat}{\lambda^2}
\nll &&  \quad
     - \Biggl[ Q_1^2                 \ln\frac{\hat{Q}^2}{m^2_1}
        + ( 1 - Q_1)^2               \ln\frac{\Shat}{m^2_2}
        +                            \ln\frac{\Shat}{m^2_l}
\nll && \quad
        - Q_1 (2 - Q_1 )             \ln y                   
        + 2 Q_1                      \ln(1 - y)            
        - 2 (1 - Q_1 + Q_1^2 )\Biggr]\ln\frac{\Shat}{2\bar\omega m_2}
\nll && \quad
    - \frac{1}{4}\Biggl[ Q_1^2       \ln^2\frac{\hat{Q}^2}{m^2_1}
        - (1 - Q_1)^2                \ln^2\frac{\Shat}{m^2_2}
        +                            \ln^2\frac{\Shat}{m^2_l}
        -2 Q_1^2 (1 - \ln y)         \ln\frac{\hat{Q}^2}{m^2_1} 
\nll && \quad
        + 2 (1 - Q_1)^2              \ln\frac{\Shat}{m^2_2}
        - 2                          \ln\frac{\Shat}{m^2_l}\Biggr]
        + \frac{1}{4} Q_1 (2 - Q_1)  \ln^2 y 
        + \frac{1}{2} Q_1            \ln^2 (1 - y)  
\nll && \quad
        - Q_1                        \ln y \ln(1 - y)                      
        + \frac{1}{2} Q_1^2          \ln y         
        - ( 1 - Q_1 + Q_1^2 )        \zeta_2
        + ( 1 - Q_1 )^2
\Biggr\}.
\ea
Soft corrections for the NC case are given by
\ba\label{SoftNC}
&& \delta^{\mathrm{Soft}}_{NC} = - \frac{\alpha}{\pi}Q_1^2
\biggl\{ \ln\frac{2\bar{\omega}}{\lambda} \biggl( 2 
- \ln\frac{\hat{Q}^2}{m_1^2} - \ln\frac{\hat{Q}^2}{m_2^2} \biggr)
\nonumber \\ && \quad
+ \frac{1}{4}\biggl(\ln\frac{\hat{Q}^2}{m_1^2} + \ln\frac{\hat{Q}^2}{m_2^2} \biggr)^2 
- \frac{1}{2}\biggl( \ln\frac{\hat{Q}^2}{m_1^2} + \ln\frac{\hat{Q}^2}{m_2^2} \biggr)
+ \zeta_2 - 1\biggr\}.
\ea
The above formula is valid both for the neutrino--quark
and antineutrino--quark NC scattering processes.

\subsection{Hard Photon Radiation}

In the case of hard photon emission, like in the process
\ba
\nu(k_1)\ +\ q_i(p_1)\ \to\ l^-(k_2)\ +\ q_f(p_2)\ +\  \gamma(k), 
\ea
we have to extend the list of kinematical variables:
\ba
\tilde{p}_2 = p_2 + k, \qquad
\tilde{Q}^2 = - 2p_1(p_2+k), \qquad
\tilde{M}^2 = - (p_2+k)^2.  
\ea
Note, that in the case without real photon radiation $\tilde{Q}^2 = \hat{Q}^2$.
Therefore in what follows we can use $\tilde{Q}^2$ instead of $\hat{Q}^2$.
The calculations of the hard photon contribution 
were evaluated also in the environment of the SANC system. 

The angular phase space of the hard photon is 
$\dd\Omega_R = \dd\cos\theta_R \dd\varphi_R$  in the $R$-system of reference,
where $\vec{k}+\vec{p}_2=0$.

For the charged current $\nu q$ scattering we have
\ba
\delta^{\mathrm{Hard}}_{\nu CC} &=& \frac{\alpha}{\pi} \biggl\{
      \Biggl[ Q_1^2                  \ln\frac{\tilde{Q}^2}{m^2_1}
        + ( 1 + Q_1)^2               \ln\frac{\hat{s}}{m^2_2}
	+       	             \ln\frac{\hat{s}}{m^2_l}
\nll &&
	+ Q_1 (2 + Q_1 )             \ln y      
	- 2 Q_1 	             \ln(1 - y)          
	- 2 (1 + Q_1 + Q_1^2 )\Biggr]\ln\frac{\hat{s}}{2\bar{\omega}m_2}
\nll &&
    -\frac{1}{2} (1+Q_1)^2                \ln^2\frac{\hat{s}}{m^2_2}
    -Q_1^2\left(\frac{17}{12}-\ln y\right)\ln\frac{\tilde{Q}^2}{m^2_1}
\nll &&
    +\frac{1}{4} (1+Q_1)^2                \ln\frac{\hat{s}}{m^2_2}
    +\frac{1}{2}\Big[2\ln y-\ln(1 - y)-y\Big]\ln\frac{\hat{s}}{m^2_l}
\nll &&
    -\left(\frac{1}{2} - Q_1 - \frac{1}{2} Q_1^2\right)\ln^2 y
    -\frac{1}{2} 			          \ln^2 (1 - y)
    +\left(\frac{1}{2}-2 Q_1 \right)	          \ln(1 - y)\ln y
\nll &&
    -\left(\frac{7}{4}-\frac{1}{2}y+\frac{3}{2}Q_1+\frac{7}{4}Q_1^2\right)\ln y
\nll &&
    +\left( 1 - y + \frac{3}{2} Q_1 \right)       \ln(1 - y)
    -\left( \frac{3}{2}+2 Q_1 \right)             \Li{2}{y}     
      -(1+Q_1)^2                                  \zeta_2
\nll &&                           
    -\frac{1}{4}\biggl[1 - 5 y - 5 Q_1 - \left(\frac{95}{18}-\frac{5}{3} y 
                      + \frac{1}{6} y^2\right) Q_1^2\biggr] \biggr\}.
\ea
Hard photon contribution to the $\bar{\nu} q$ CC scattering reads
\ba
\delta^{\mathrm{Hard}}_{\bar{\nu} CC} &=& \frac{1}{(1-y)^2}\,\frac{\alpha}{\pi} 
\biggl\{
   (1-y)^2 \Biggl[ Q_1^2            \ln\frac{\tilde{Q}^2}{m^2_1}
        + ( 1 - Q_1)^2              \ln\frac{\hat{s}}{m^2_2}
	+       	            \ln\frac{\hat{s}}{m^2_l}
\nll &&
	+ Q_1 (Q_1 - 2)             \ln y      
	+ 2 Q_1 	             \ln(1 - y)          
	- 2 (1 - Q_1 + Q_1^2 )\Biggr]\ln\frac{\hat{s}}{2\bar{\omega}m_2}
\nll &&
    -\frac{1}{2} (1-Q_1)^2 (1-y)^2 \ln^2\frac{\hat{s}}{m^2_2}
    +\frac{1}{4} (1-Q_1)^2 (1-y)^2 \ln\frac{\hat{s}}{m^2_2}
\nll &&
    -Q_1^2(1-y)^2\left(\frac{17}{12}-\ln y\right)\ln\frac{\tilde{Q}^2}{m^2_1}
    + \Big[(1-y)^2\ln\frac{y}{1-y} + y - \frac{3}{4}y^2\Big]\ln\frac{\hat{s}}{m^2_l}
\nll &&
    - \frac{1}{2}(1-y)^2\left(1 + 2Q_1 - Q_1^2\right)\ln^2 y
    - (1-y)^2 \ln^2 (1 - y)
\nll &&
    + (1-y)^2\left(1+ 2Q_1 \right)\ln(1 - y)\ln y
    + \biggl( - \frac{7}{4} + \frac{5}{2}y 
       + \frac{3}{2}(1-y)^2Q_1 - y^2 
\nll &&
       - \frac{7}{4}(1-y)^2Q_1^2 \biggr)\ln y
    + \left( \frac{7}{4} - \frac{5}{2}y 
      + y^2 - \frac{3}{2}(1-y)^2Q_1 \right) \ln(1 - y)
\nll &&                           
    - (1-y)^2 \left(1 - 2 Q_1 \right) \Li{2}{y}     
      -(1-y)^2(1-Q_1)^2  \zeta_2 
   - \frac{1}{4} + yQ_1 
\nll &&                           
    - \frac{20}{9}yQ_1^2 - y + \frac{17}{18}y^2Q_1^2 
    + y^2 - \frac{5}{4}Q_1 + \frac{95}{72}Q_1^2 \biggr\}.
\ea

For the neutral current scattering we have
\ba \label{HardNC}
\delta^{\mathrm{Hard}}_{\nu NC} &=& 
\frac{\alpha}{\pi}Q_1^2 \biggl\{ 
\ln\frac{\Shat}{2\bar{\omega}m_2} \biggl( \ln\frac{\Shat}{m_2^2}
+ \ln\frac{\tilde{Q}^2}{m_1^2} + \ln y -2 \biggr)
+ \ln y \ln\frac{\tilde{Q}^2}{m_1^2}
\nonumber \\
&-& \frac{17}{12}\ln\frac{\tilde{Q}^2}{m_1^2} 
- \frac{1}{2}\ln^2\frac{\Shat}{m_2^2}
+ \frac{1}{4}\ln\frac{\Shat}{m_2^2} 
+ \frac{1}{2}\ln^2y - \zeta_2 - \frac{7}{4}\ln y 
\nonumber \\
&+& \frac{17}{18} 
+ \frac{1}{g_L^2+g_R^2(1-y)^2}\biggl( \frac{1}{3}(1-y)(g_L^2+g_R^2)
+ \frac{1}{24}(g_L^2(1-y)^2+g_R^2) \biggr)
\biggr\}.
\ea
The corresponding correction to the NC antineutrino--quark scattering, 
$\delta^{\mathrm{Hard}}_{\bar{\nu} NC}$, can be obtained from the above
equation by the substitution $g_L^2 \leftrightarrow g_R^2$.

\subsection{Quark Mass Singularities}

In the sum of the soft and hard photonic corrections, the auxiliary
parameter $\bar\omega$ (soft-hard separator) cancels out. The infrared
singular terms (with logarithm of $\lambda)$ cancel out in the sum of 
the virtual and soft contributions.
Moreover in the total sum, the large logarithms (mass singularities)
with the final state quark mass $m_2$ disappear in accordance with 
the Kinoshita--Lee-Nauenberg
theorem~\cite{Kinoshita:1962ur,Lee:1964is}. 

Large logarithms containing the initial quark mass, $\ln(\hat{Q}^2/m_1^2)$,
remain in the sum of all contributions. But these logs have been
already effectively taken into account in the parton density functions (PDFs).
In fact, QED radiative corrections to the quark line are 
usually not taken into account in procedures of PDF extraction.
Moreover, the leading log behaviors of the QED and QCD DGLAP evolution of PDFs in
the leading order are proportional to each other. So one gets evolution
of PDFs with an effective coupling constant 
\ba
\alpha^{\mathrm{eff}}_{s} \approx \alpha_{s} + \frac{Q_1^2}{C_F}\alpha,
\ea
where $\alpha_s$ is the strong coupling constant, and $C_F$ is the QCD color factor.
The nontrivial difference between the QED evolution and the QCD one starts to appear 
from higher orders,
and the corresponding numerical effect is small compared to the remaining QCD
uncertainties in PDFs~\cite{Kripfganz:bd,Spiesberger:1994dm,Roth:2004ti}.
See also Ref.~\cite{Diener:2003ss} for a discussion with respect to the process 
under consideration. 

The best approach to the whole problem would be to re--analyze  all
the experimental DIS data taking into account QED corrections to the quark line
at least at the next--to--leading order. But for the present task we can limit
ourselves with application of the \MSbar subtraction scheme~\cite{Bardeen:1978yd} 
to the QED part of the radiative corrections for the process under consideration. 
This leads to a shift of the
initial quark mass singularities (with a certain constant term) from our
result into the corresponding quark density function. 
The latter should be also taken in the \MSbar scheme\footnote{A similar consideration
can be performed in the DIS and other schemes as well.}. 
In fact, using the initial condition for the non--singlet NLO QED quark 
structure function
(which coincides with the QCD one with the trivial substitution 
$C_F\alpha_s\to Q_1^2\alpha$, see Ref.~\cite{Berends:1987ab}), 
we get the following expression for the terms to be subtracted from the full 
calculation with massive quarks:
\ba
\delta_{\msbar}  &=& Q_1^2 \frac{\alpha}{2\pi} \int\limits_0^1 \dd x\; x \biggl[
\frac{1+x^2}{1-x} \biggl( \ln\frac{\hat{Q}^2}{m_1^2}  - 1 - 2\ln(1-x) 
\biggr) \biggr]_+ 
\nonumber \\
&=& Q_1^2 \frac{\alpha}{2\pi}\biggl( - \frac{4}{3} \ln\frac{\hat{Q}^2}{m_1^2} 
- \frac{17}{9} \biggr). 
\ea

\subsection{The Muon Mass Singularity} 

In the case of CC scattering, the large logarithms singular in
the limit $m_l \to 0$ remain in the final answer. These terms are 
in agreement with the prediction of the renormalization group approach.
In fact, they can be described by the electron (muon)
fragmentation (structure) function 
approach~\cite{Kuraev:hb,Skrzypek:1992vk,Arbuzov:1999cq}:
\ba
\dd\sigma^{LL} = \sum_j\dd\tilde{\sigma}_j \otimes D_{\mu j}(\xi,Q^2),
\ea
where $\dd\tilde{\sigma}_j$ is the differential cross section of 
neutrino DIS with production of particle $j$ ($j=\mu,\ \gamma$, etc.);  
$D(\xi,Q^2)$ describes the probability to find a muon with the relative
energy fraction $\xi$ in particle $j$; sign $\otimes$ stands for 
the convolution operation.

For our purposes it is enough to consider only the leading log approximation
in the $\order{\alpha L}$ and $\order{\alpha^2L^2}$ orders including the
contribution of electron--positron pairs.
Under these conditions we can write the QED fragmentation function
in the form:
\ba \label{Dfun}
D^{NS}_{\mu\mu}(\xi,Q^2) &=& \delta(1-\xi) + \frac{\alpha}{2\pi}LP^{(0)}(\xi)
+ \frac{1}{2}\left(\frac{\alpha}{2\pi}L\right)^2 P^{(0)}(\xi)\otimes P^{(0)}(\xi) 
\nonumber \\
&+& \frac{1}{3}\left(\frac{\alpha}{2\pi}L\right)^2 P^{(0)}(\xi) 
+ \order{\alpha,\alpha^2L,\alpha^3L^3},
\ea
where $L$ is the so--called large logarithm, 
and $P^{(0)}(\xi)$ is the lowest order non--singlet splitting function:
\ba
P^{(0)}(\xi) = \left[\frac{1+\xi^2}{1-\xi}\right]_+. 
\ea
For the definition of the plus prescription and the convolution operation
see for instance Refs.~\cite{Altarelli:1981ax,Arbuzov:1999cq}.

Differential cross section $\dd\hat{\sigma}_j$ (for $j=\mu$) is the 
Born level distributions~(\ref{SnuqCC}) with a shifted value of variable $y$:
\ba
\tilde{y} = \frac{\xi + y - 1}{\xi},
\ea
which provides the proper value of variable $y$ after the fragmentation stage under
the adopted condition for particle registration.
Convolution with the $P^{(0)}$ splitting function gives the following first order
leading logarithmic corrections:
\ba
&& \delta_{\nu CC}^{(1)LL} =  \frac{\alpha}{2\pi}L\biggl(2\ln y 
- \ln(1-y) + \frac{3}{2} - y \biggr),
\\
&& \delta_{\bar{\nu} CC}^{(1)LL} =  \frac{1}{(1-y)^2}\,\frac{\alpha}{2\pi}L\biggl(
2(1-y)^2\ln\frac{y}{1-y} + \frac{3}{2} - y \biggr).
\ea
We will see that the above expressions reproduce the the leading log part of
the complete result for the $\order{\alpha}$ correction 
in Eqs.~(\ref{CCRCnu},\ref{CCRCnb}).

By convolution with the $\order{\alpha^2}$ term from Eq.~(\ref{Dfun}), we
get the second order leading log corrections to the CC neutrino DIS:
\ba
&& \delta_{\nu CC}^{(2)LL} =  \frac{1}{2}
\left(\frac{\alpha}{2\pi}L\right)^2 \biggl[
4\ln^2y - 4\ln y\ln(1-y) + \frac{1}{2}\ln^2(1-y) - 4\zeta_2
\nonumber \\ && \quad
+ (6-4y)\ln y + (3y-4)\ln(1-y) + \frac{9}{4} \biggr] 
+ \frac{1}{3}\,\frac{\alpha}{2\pi}L \delta_{\nu CC}^{(1)LL},
\\
&& \delta_{\bar{\nu} CC}^{(2)LL} = \frac{1}{(1-y)^2}\,\frac{1}{2}
\left(\frac{\alpha}{2\pi}L\right)^2 \biggl[
4(1-y)^2\biggl( \Li{2}{1-y} + \ln^2y + \frac{1}{2}\ln^2(1-y) 
\nonumber \\ && \quad
- 2\zeta_2\biggr)+ (6-4y)\ln y + \biggl(y-\frac{3}{2}\biggr)\ln(1-y) 
+ \frac{9}{4} - 2y \biggr] 
+ \frac{1}{3}\,\frac{\alpha}{2\pi}L \delta_{\bar{\nu} CC}^{(1)LL}.
\ea
Argument of the large logarithm depends on the energy scale of the process
under consideration and the muon mass. We have two scales: $\hat s$ and
$\hat{Q}^2$. In the actual calculations the logarithm with the muon mass 
singularity arises in the form $L=\ln(\hat{s}/m_l^2)$. So we choose $\hat{s}$,
while the difference between the two possibilities appears in our case only 
in terms of the order $\order{\alpha^2L}$, which are omitted now in any case.

Using the formalism of the fragmentation function approach~\cite{Mele:1990cw}
we can get the next--to--leading corrections\footnote{See Ref.~\cite{Arbuzov:2002cn} 
for an analogous calculation.} of the order $\order{\alpha^2L}$.
But, as will be seen from our numerical estimates below, they are small compared 
with  the requested precision tag.

\section{Numerical Results and Conclusions
\label{NumRes}}

Summing up all the different RC contributions considered above, and applying the
\MSbar subtraction of the initial state quark mass singularity we arrive
at the result for the corrections to neutrino--quark cross sections:
\ba
\frac{\dd^2\sigma_i^{\mathrm{Corr.}}}{\dd x\; \dd y} = 
\frac{\dd^2\sigma_i^{\mathrm{Born}}}{\dd x\; \dd y}
(1 + \delta_i + \delta_{i}^{(2)LL} - \delta_{\msbar}),
\ea
where $\delta_{i}^{(2)LL}$ vanishes in the NC case.

For neutrino--quark CC scattering we have
\ba \label{CCRCnu}
\delta_{\nu CC} &=& \frac{\alpha}{\pi}\biggl\{
- \frac{3}{2}\ln\frac{\hat{s}}{M_{\sss Z}^2} 
+ \biggl( \frac{3}{4} - \frac{1}{2}y -\frac{1}{2}\ln(1 - y) 
+ \ln y\biggr)\ln\frac{\hat s}{m_l^2}
\nonumber \\
&+& \frac{1}{2}\ln(1-y)\ln y
- \frac{3}{2}\Li{2}{y} - \frac{1}{2}\ln^2y - \frac{1}{2}\ln^2(1-y)
\nonumber \\
&+& (1-y)\ln(1-y) - \frac{7}{4}\ln y + \frac{1}{2}y\ln y
+ \frac{5}{4}y + \frac{1}{2} + 2\zeta_2 
\nonumber \\
&+& Q_1\biggl( 3 -\frac{3}{2}\ln\frac{\hat{s}}{M_{\sss Z}^2} + \zeta_2 
- \ln(1-y)\ln y - 2\Li{2}{y}\biggr)
\nonumber \\
&+& Q_1^2\biggl( -\frac{2}{3}\ln\frac{\hat{Q}^2}{m_1^2}
+ \frac{23}{72} - \frac{5}{12}y + \frac{1}{24}y^2 - \zeta_2 \biggr)
\biggr\},
\ea
where we used the charge conservation law $Q_1=Q_2+Q_l$
and the explicit value $Q_l=-1$.
For antineutrino--quark CC scattering we have
\ba \label{CCRCnb}
\delta_{\bar{\nu}\; CC} &=& \frac{1}{(1-y)^2}\,\frac{\alpha}{\pi}\biggl\{
\biggl( (1-y)^2\ln\frac{y}{1-y} - \frac{y}{2} 
+ \frac{3}{4} \biggr)\ln\frac{\hat s}{m_l^2}
+ (1-y)^2\biggl( 2\zeta_2 - \Li{2}{y}
\nonumber \\
&+& \ln(1-y)\ln y - \frac{1}{2}\ln^2y 
- \ln^2(1-y)  \biggr)
+ \biggl( -\frac{7}{4} + \frac{5}{2}y - y^2\biggr)\ln\frac{y}{1-y}
\nonumber \\
&-& \frac{5}{4}  + y 
+ Q_1\biggl[ \frac{1}{2} - \frac{5}{2}y + \frac{7}{4}y^2 
- \frac{3}{2}(1-y)^2 \ln\frac{\hat{s}(1-y)}{M_{\sss Z}^2} 
\nonumber \\
&+& (1-y)^2\biggl(2\Li{2}{y} +\ln(1-y)\ln y - \zeta_2 \biggr)\biggr]
+ Q_1^2\biggl[ - \frac{2}{3}(1-y)^2\ln\frac{\hat{Q}^2}{m_1^2}
\nonumber \\
&-& (1-y)^2\zeta_2 - \frac{1}{18}y^2 - \frac{2}{9}y + \frac{23}{72} \biggr]
\biggr\}.
\ea

For the case of NC neutrino scattering we get
\ba \label{NCRCnu}
\delta_{\nu NC} &=& \frac{1}{g_L^2 + g_R^2(1-y)^2}
\biggl\{ \frac{\alpha}{\pi}Q_1^2\biggl[
 g_R^2  \biggl(  - \frac{2}{3}(1-y)^2\ln\frac{\hat{Q}^2}{m_1^2} 
- \frac{1}{18}(1-y)^2 + \frac{1}{3}(1-y) 
\nonumber \\
&+& \frac{1}{24} -  \zeta_2(1-y)^2 \biggr)
+ g_L^2 \biggl(  - \frac{2}{3}\ln\frac{\hat{Q}^2}{m_1^2}
+ \frac{1}{24}(1-y)^2 + \frac{1}{3}(1-y) 
\nonumber \\
&-&  \frac{1}{18} - \zeta_2 \biggr)
\biggr] + \tilde{g}_L^2 + \tilde{g}_R^2(1-y)^2
\biggr\} - 1.
\ea
The expression for the one--loop radiative correction to the 
antineutrino NC process can be received from the above one using the substitution:
\ba \label{NCRCnb}
\delta_{\bar{\nu}NC} &=& \delta_{\nu NC} (g_L \leftrightarrow g_R,\ \ 
\tilde{g}_L \leftrightarrow \tilde{g}_R).
\ea

Formulae (\ref{CCRCnu},\ref{CCRCnb},\ref{NCRCnu}) completely agree with the ones 
derived in Ref.~\cite{Bardin:1986bc}. Moreover, our formulae for the CC scattering
case agree with the calculations presented in Ref.~\cite{Wheater:yk}, where
the explicit expressions were given for $\nu d$ and $\bar{\nu} u$ CC scattering
channels. For the comparison, one should subtract from their formulae for
$g^\nu(y,S)$ and $g^{\bar{\nu}}(y,S)$ the quantities $C$ and $(1-y)^2C$, 
respectively,
\ba
C = \ln\frac{M_{\sss W}^3}{\mu M_{\sss Z}^2} + \frac{1}{4}.
\ea  
The subtraction reflects the choice of the \MSbar renormalization scheme.

Let us consider numerical results obtained for the following conditions: 
the fixed neutrino energy $E_\nu = 80$~GeV; isoscalar nuclear target;
cut on the energy of the final state hadronic system 
$\hat{E}_{\mathrm{hadr}} \geq 10$~GeV. We used the CTEQ4L set~\cite{Lai:1996mg} 
of parton density functions. In Table~1 we present
numerical values and the absolute shifts due to radiative corrections of
the quantities~\cite{Paschos:1972kj}, 
constructed from the cross sections of neutrino DIS,
\ba
&& R^\nu = \frac{\sigma^{\nu}_{NC}(\nu_\mu N\to\nu_\mu X)}
{\sigma^{\nu}_{CC}(\nu_\mu N\to\mu^- X)},
\\ \label{PW}
&& R^- = \frac{\sigma^{\nu}_{NC}(\nu_\mu N\to\nu_\mu X)
- \sigma^{\bar{\nu}}_{NC}(\bar{\nu}_\mu N\to\bar{\nu}_\mu X)}
{\sigma^{\nu}_{CC}(\nu_\mu N\to\mu^- X)
- \sigma^{\bar{\nu}}_{CC}(\bar{\nu}_\mu N\to\mu^+ X)},
\\
&& \delta R^\nu_{NC} = \frac{\sigma^{\mathrm{Corr.}}_{\nu NC}
-\sigma^{\mathrm{Born}}_{\nu NC}}{\sigma^{\mathrm{Born}}_{\nu NC}}\, ,
\qquad 
\delta R^\nu_{CC} = - \frac{\sigma^{\mathrm{Corr.}}_{\nu CC}
-\sigma^{\mathrm{Born}}_{\nu CC}}{\sigma^{\mathrm{Born}}_{\nu CC}}\, .
\ea
The Born level values are $R_0^{\nu}=0.31764$ and $R_0^{-} = 0.27831$.
In the computations we used the following values of constants and parameters 
taken from \cite{Hagiwara:fs} (the same as the ones given by Eq.~(4.6) 
of Ref.~\cite{Diener:2003ss}). Flag {\tt ISCH} defines the choice of
the initial quark mass singularity treatment: {\tt ISCH}=0 means no any subtraction
(as has been done in Ref.~\cite{Bardin:1986bc}), and {\tt ISCH}=1 corresponds to the
\MSbar subtraction scheme as discussed above. The order of the series in the leading 
large logarithms with the lepton mass singularity is governed by flag {\tt ILLA}:
$O(\alpha L)$ for {\tt ILLA}=1 and $O(\alpha^2 L^2)$ for {\tt ILLA}=2.

\begin{table} 
\begin{tabular}{|c|c|c|c|c||c|c|}
\hline 
({\tt ISCH,ILLA}), & 
$R^{\nu}$ & 
$\delta R^{\nu}_{NC}$ & 
$\delta R^{\nu}_{CC}$ & 
$\Delta^\nu \sww$ & 
$R^{-}$ & 
$\Delta^- \sww$
 \\ 
 EW scheme &  &  &  &  &  & 
 \\ \hline
(0,1), $G_F$ & 0.31006 & $-$0.00291 & $-$0.02147 & $-$0.01130 & 0.27094 & $-$0.00737 \\
(1,1), $G_F$ & 0.31063 & ~~ 0.00071 & $-$0.02327 & $-$0.01044 & 0.27195 & $-$0.00636 \\
(1,2), $G_F$ & 0.31067 & ~~ 0.00071 & $-$0.02315 & $-$0.01039 & 0.27196 & $-$0.00637 \\
\hline
(1,2), $\alpha(0)$ & 
               0.31080 & $-$0.05816 & ~~ 0.03743 & $-$0.01020 & 0.27209 & $-$0.00623 \\
\hline
\end{tabular}
\caption{Effect of RC on $R^\nu$, $R^-$, and $\sww$ values in different 
approximations.}
\end{table}

For the illustrations we used the simple tree--level relations between the 
shifts of $R^{\nu,-}$ and $\sww$:
\ba
\delta R^\nu = \biggl(1 - \frac{40}{27}\sww\biggr)\Delta^{\nu}\sww, \qquad
\delta R^- = \Delta^{-}\sww,
\ea
where $\delta R^\nu$ and $\delta R^-$ are the differences between the
corrected quantities and the tree--level ones.
In a real case of experimental data analysis the contribution of 
radiative corrections should be estimated within 
a general fit including all other effects like detector efficiencies, 
nuclear shadowing {\it etc.} 

One can see by comparing the first two lines in Table~1, 
that the \MSbar subtraction of the initial quark mass singularity
gives a numerically important contribution. We considered two electroweak schemes:
the $G_{\mathrm{Fermi}}$ and $\alpha(0)$ (see Ref.~\cite{Diener:2003ss} 
and references therein for the definition). 
We found that the EW scheme dependence is visible, but
less than the one observed in Ref.~\cite{Diener:2003ss}. 

One can compare our numbers with the corresponding results given in 
Table~1 of Ref.~\cite{Diener:2003ss} for the case of the energy cut 
$E_{\mathrm{had+phot}}^{\mathrm{LAB}}>10$~GeV. The difference between the
two calculations appeared to be not small. Partially this is related to 
different treatments of the effect of the scaling violation in the parton
density functions. A more detailed comparison is in progress and will be 
published elsewhere.

In Figures~1,2 we present the relative value of the sum of 
radiative corrections,
\ba \nonumber 
\delta_i + \delta_{i}^{(2)LL} - \delta_{\msbar}),  
\ea
for two channels (neutrino$-$up quark NC scattering and
neutrino$-$down quark CC scattering) as a function of $y$ for three
fixed values of $x$. The neutrino beam energy and other parameters are
the same as the ones defined for Table~1.
One can see that the behavior of the corrections
is rather smooth. Note that in the case of a different choice of
variables, when $Q^2$ is defined from the observation of the outgoing
muon, the size of the corrections for the CC case becomes several times
larger\footnote{The same effect is well known in the conventional charged
lepton deep inelastic scattering.}.

\begin{figure}[ht]
\begin{center}
\includegraphics*[width=12cm,height=5cm,angle=0]{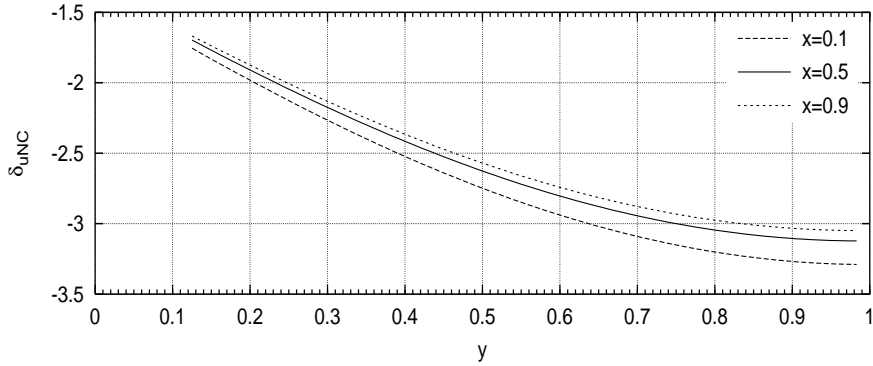}
\end{center}
\caption{Relative effect of radiative corrections to $\nu-u$ NC scattering
as a function of $y$ for three fixed values of $x$.}
\label{Fig1}
\end{figure}

\begin{figure}[ht]
\begin{center}
\includegraphics*[width=12cm,height=5cm,angle=0]{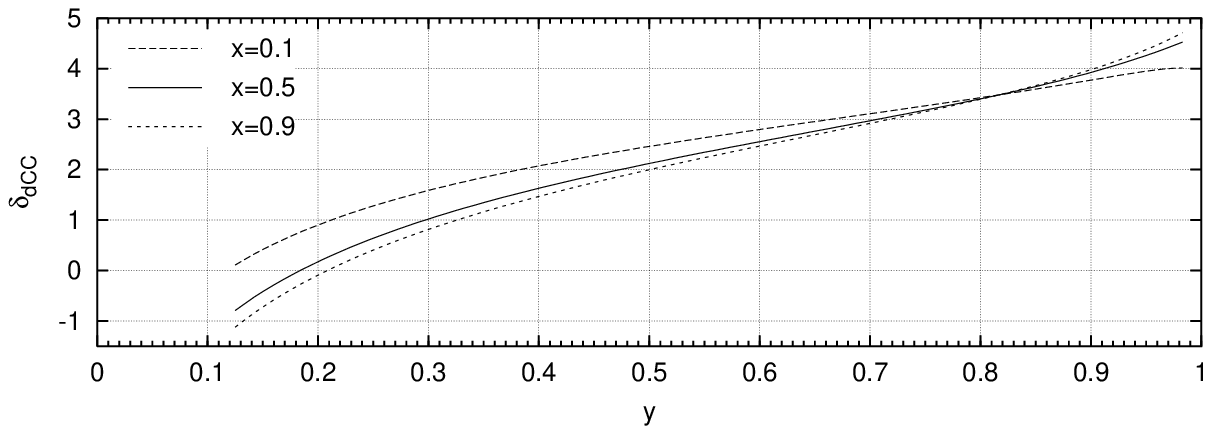}
\end{center}
\caption{Relative effect of radiative corrections to $\nu-d$ CC scattering
as a function of $y$ for three fixed values of $x$.}
\label{Fig2}
\end{figure}

Let us consider the sources of the theoretical uncertainty, related to
the incomplete knowledge of the radiative corrections. The sources are
listed in Table~2.
\begin{table} 
\begin{tabular}{|c|c|c|c|}
\hline 
Source of uncertainty   & \multicolumn{3}{c|}{Estimated value} \\ 
\hline
                        & for $R^\nu$       & for $\Delta^{\nu}\sww$& for $R^-$ \\
\hline
higher order EW      RC & $1.5\cdot10^{-4}$ & $2.2\cdot10^{-4}$     & $1.2\cdot10^{-4}$ \\  
higher order QED     RC & $0.5\cdot10^{-4}$ & $0.7\cdot10^{-4}$     & $0.2\cdot10^{-4}$ \\  
$\order{\alpha\alpha_s}$& $2.2\cdot10^{-4}$ & $3.3\cdot10^{-4}$     & $1.5\cdot10^{-4}$ \\  
\hline
all combined            & $2.7\cdot10^{-4}$ & $4.0\cdot10^{-4}$     & $2.0\cdot10^{-4}$ \\  
\hline
\end{tabular}
\caption{Estimates of different contributions to the theoretical uncertainty.}
\end{table}
First, we have observed already the electroweak scheme dependence. From the
general point of view, the $G_{\mathrm{Fermi}}$ scheme looks preferable for
our problem. By trying the other EW scheme, we get an estimate of how large
can be unknown higher order electroweak corrections. The pure QED 
corrections\footnote{Unambiguous separation of EW and QED corrections in
the CC channel is not possible.}
are in our case not large, and by looking at the leading second order logarithmic
contribution, we can put a limit on unknown higher order QED contributions.
As the main source of the theoretical uncertainty we consider the contribution 
of the radiative corrections of the order $\order{\alpha\alpha_s}$. They can appear
as in the loop insertions into the $W$-boson propagator as well as in amplitudes
where photonic and gluonic lines appear simultaneously. In our approach, we separated
the EW RC from the QCD ones. After all, we receive a direct product of QCD
effects (see i.e. Ref.~\cite{Alekhin:2002fv}) and EW corrections, which doesn't
cover the full $\order{\alpha\alpha_s}$ answer. Moreover, as can be seen
from our calculations, we always consider $\tilde{Q}^2$ to be the argument of
the partonic density function. But in the case of hard photon radiation off
lepton, $\tilde{Q}^2$ doesn't coincide with the square of the ``true'' hadronic
momentum transfer. This leads to a certain effect in $\order{\alpha\alpha_s}$.
To estimate the corresponding uncertainty we varied the value of $Q^2$ in 
the argument of the PDFs. 
As the result of our estimates we found that our result for radiative corrections
can receive up to $\pm 3\%$ of a relative shift. Having a unique value of 
uncertainty for the functions of two variables seems to be reasonable, because
the functions are relatively flat, and because, in the derivation of the
number we looked both at the variations at the differential level and at the
ones for the integrated cross sections. Note that we estimated the uncertainties
in the form of maximal variations.  
A work is going on to extend our consideration to the case of other
variables and to avoid the effect of improper argument in PDFs, mentioned above.

It is worth to note, that radiative corrections to the same series of processes
calculated in a different set of variables can have a completely different
behavior. For example, in the variables adopted by the NuTeV experiment,
the size of corrections to the CC scattering (where the energy and angle 
of the outgoing charge lepton can be measured) are much bigger than the
corrections presented here. In the same manner, estimates of the theoretical
uncertainties should be considered taking into account the choice of variables
and, may be, other relevant experimental conditions.
Nevertheless we agree with the 
authors of Ref.~\cite{Diener:2003ss}, who claimed that the higher order
contributions to the theoretical uncertainty seem to have been underestimated 
by the NuTeV experiment~\cite{McFarland:gx}. 

Our analytical results are combined into a {\tt FORTRAN} code, which
can be directly used for experimental data analysis.
As can be seen from the numerical estimates the effect of radiative
corrections is greater than the present experimental uncertainty, see {\it e.g.}
the NuTeV~\cite{McFarland:gx} result:
\ba
\sww = 0.2277 \pm 0.0013(\mathrm{stat.}) \pm 0.0009(syst.).
\ea
That makes it important to take RC into account in the proper way.
Estimates of the theoretical uncertainty becomes also relevant for
the derivation of the resulting error in the analysis of experimental 
data in the present and future precision experiments on neutrino DIS.

\ack{
We are grateful to K.~Diener, S.~Dittmaier, W.~Hollik, and R.~Petti
for discussions.
This work was supported by the INTAS grant 03-51-4007.
One of us (A.A.) thanks the RFBR grant 03-02-17077.
}

\section*{Appendix A\\
List of Integrals for Soft Photon RC
}
\setcounter{equation}{0}
\renewcommand{\theequation}{A.\arabic{equation}}

All relevant integrals are defined in the reference frame $\vec{p}_2 = 0$.
They are:
\ba
&& \int\frac{\dd^3k}{\omega} \frac{p_1^2}{(p_1k)^2} = 
4\pi\left(\ln\frac{2\bar{\omega}}{\lambda} - \frac{1}{2}L_1 \right),
\\
&& \int\frac{\dd^3k}{\omega} \frac{p_2^2}{(p_2k)^2} = 
4\pi\left(\ln\frac{2\bar{\omega}}{\lambda} - 1 \right),
\\
&& \int\frac{\dd^3k}{\omega} \frac{k_2^2}{(k_2k)^2} = 
4\pi\left(\ln\frac{2\bar{\omega}}{\lambda} - \frac{1}{2}L_l \right),
\\
&& \int\frac{\dd^3k}{\omega} \frac{2p_1p_2}{(p_1k)(p_2k)} = 
2\pi\left( 2\ln\frac{2\bar{\omega}}{\lambda}L_1 - \frac{1}{2}L_1^2
- 2\zeta(2) \right),
\\
&& \int\frac{\dd^3k}{\omega} \frac{2p_2k_2}{(p_2k)(k_2k)} = 
2\pi\left(2\ln\frac{2\bar{\omega}}{\lambda}L_l - \frac{1}{2}L_l^2
- 2\zeta(2) \right),
\\
&& \int\frac{\dd^3k}{\omega} \frac{2p_1k_2}{(p_1k)(k_2k)} = 
2\pi\left[ 2 \ln\frac{2\bar{\omega}}{\lambda}\left(
\ln\frac{w}{m_l^2} + \ln\frac{w}{m_1^2} \right)
- \ln\frac{w}{m_1^2}L_1 - \ln\frac{w}{m_l^2}L_l 
\right. \nonumber \\ && \left.
+ \frac{1}{2}\ln^2\frac{w}{m_1^2}
+ \frac{1}{2}\ln^2\frac{w}{m_l^2}
- \ln^2\frac{\hat{Q}^2}{\hat{s}} - 4\zeta(2) 
+ 2\Li{2}{\frac{1+c}{2}} 
\right],
\\ \nonumber
&& L_1 = \ln\frac{4(p_1^0)^2}{m_1^2}=\ln\frac{\hat{Q}^2}{m_1^2}
+ \ln\frac{\hat{Q}^2}{m_2^2}, \qquad
L_l = \ln\frac{4(k_2^0)^2}{m_l^2} = \ln\frac{\hat{s}^2}{m_l^2m_2^2},
\\ \nonumber
&& w = 2p_1k_2 \approx \hat{s} - \hat{Q}^2, \qquad \hat{Q}^2 = 2m_2p_1^0,
\\ \nonumber
&& \frac{1+c}{2} \equiv \frac{1+\cos(\widehat{\vec{p}_1\vec{k}}_2)}{2} =
1 - \frac{m_2^2(\hat{s}-\hat{Q}^2)}{\hat{s}\hat{Q}^2} \approx 1.
\ea
Terms which are suppressed by the ratio of a quark or muon mass to
a large energy scale (beam energy or momentum transferred) are
neglected.


\begin{thebibliography}{99}

\bibitem{Astier:2003rj}
P.~Astier {\it et al.}  [NOMAD Collaboration],
Nucl.\ Instrum.\ Meth.\ A {\bf 515} (2003) 800.

\bibitem{McFarland:gx}
K.~S.~McFarland {\it et al.},
Int.\ J.\ Mod.\ Phys.\ A {\bf 18} (2003) 3841; \\
G.P. Zeller {\it et al.} [NuTeV Collaboration],
Phys.\ Rev.\ Lett. {\bf 88} (2002) 091802
[Erratum-ibid. {\bf 90} (2003) 239902].

\bibitem{Kayis-Topaksu:ds}
A.~Kayis-Topaksu {\it et al.}  [CHORUS Collaboration],
Phys.\ Lett.\ B {\bf 575}, 198 (2003).

\bibitem{Bardin:1986bc}
D.~Y.~Bardin and V.~A.~Dokuchaeva,
{\em On The Radiative Corrections To The Neutrino Deep Inelastic Scattering,\/}
Preprint JINR-E2-86-260, Dubna, 1986.

\bibitem{Andonov:2002jg}
A.~Andonov, D.~Bardin, S.~Bondarenko, 
P.~Christova, L.~Kalinovskaya, G.~Nanava and G.~Passarino,
{\em Project SANC (former CalcPHEP): Support of analytic and numeric calculations
for experiments at colliders,\/}
Published in in proceedings of ICHEP, Amsterdam, The Netherlands, July 2002,
p.825 [hep-ph/0209297].

\bibitem{Bardin:zd}
D.~Bardin, P.~Christova and L.~Kalinovskaya,
Nucl.\ Phys.\ Proc.\ Suppl.\  {\bf 116} (2003) 48.

\bibitem{SANCwww}
SANC project website: http://brg.jinr.ru

\bibitem{Andonov:2002xc}
A.~Andonov, D.~Bardin, S.~Bondarenko, P.~Christova, L.~Kalinovskaya and G.~Nanava,
Phys.\ Part.\ Nucl.\  {\bf 34} (2003) 577.

\bibitem{BP:book}
D. Bardin, G. Passarino,
{\em The Standard Model in the Making: Precision Study
of Electroweak Interactions,\/} Oxford, Clarendon, 1999.

\bibitem{Kinoshita:1962ur}
T.~Kinoshita,
J.\ Math.\ Phys.\  {\bf 3} (1962) 650.

\bibitem{Lee:1964is}
T.~D.~Lee and M.~Nauenberg,
Phys.\ Rev.\  {\bf 133} (1964) B1549.

\bibitem{Kripfganz:bd}
J.~Kripfganz and H.~Perlt,
Z.\ Phys.\ C {\bf 41} (1988) 319.

\bibitem{Spiesberger:1994dm}
H.~Spiesberger,
Phys.\ Rev.\ D {\bf 52} (1995) 4936.

\bibitem{Roth:2004ti}
M.~Roth and S.~Weinzierl,
hep-ph/0403200.

\bibitem{Diener:2003ss}
K.~P.~O.~Diener, S.~Dittmaier and W.~Hollik,
Phys.\ Rev.\ D {\bf 69} (2004) 073005.


\bibitem{Bardeen:1978yd}
W.~A.~Bardeen, A.~J.~Buras, D.~W.~Duke, and T.~Muta,
Phys.\ Rev.\ D {\bf 18} (1978) 3998.

\bibitem{Berends:1987ab}
F.~A.~Berends, W.~L.~van Neerven, and G.~J.~Burgers,
Nucl.\ Phys.\ B {\bf 297} (1988) 429,
{\it Erratum: ibid.} B {\bf 304} (1988) 921.

\bibitem{Kuraev:hb}
E.~A.~Kuraev and V.~S.~Fadin,
Sov.\ J.\ Nucl.\ Phys.\  {\bf 41} (1985) 466.

\bibitem{Skrzypek:1992vk}
M.~Skrzypek,
Acta Phys.\ Polon.\ B {\bf 23}, 135 (1992).

\bibitem{Arbuzov:1999cq}
A.B. Arbuzov, 
Phys.\ Lett.\ B {\bf 470} (1999) 252.

\bibitem{Altarelli:1981ax}
G.~Altarelli,
Phys.\ Rept.\  {\bf 81} (1982) 1.


\bibitem{Mele:1990cw}
B.~Mele and P.~Nason,
Nucl.\ Phys.\ B {\bf 361} (1991) 626.

\bibitem{Arbuzov:2002cn}
A.~Arbuzov and K.~Melnikov,
Phys.\ Rev.\ D {\bf 66} (2002) 093003.

\bibitem{Wheater:yk}
J.~F.~Wheater and C.~H.~Llewellyn Smith,
Nucl.\ Phys.\ B {\bf 208} (1982) 27
[Erratum-ibid.\ B {\bf 226} (1983) 547].

\bibitem{Lai:1996mg}
H.~L.~Lai {\it et al.},
Phys.\ Rev.\ D {\bf 55} (1997) 1280.

\bibitem{Paschos:1972kj}
E.~A.~Paschos and L.~Wolfenstein,
Phys.\ Rev.\ D {\bf 7} (1973) 91.

\bibitem{Hagiwara:fs}
K.~Hagiwara {\it et al.}  [Particle Data Group Collaboration],
Phys.\ Rev.\ D {\bf 66} (2002) 010001.

\bibitem{Alekhin:2002fv}
S.~Alekhin,
Phys.\ Rev.\ D {\bf 68} (2003) 014002.

\end{thebibliography}
\end{document}